\documentstyle[a4wide]{article}
\title{INTRINSIC AND OPERATIONAL OBSERVABLES IN QUANTUM MECHANICS\thanks{
Presented at the Fourth Central--European Workshop on Quantum Optics,
Budmerice, Slovakia, May 31 -- June 3, 1996}}
\author{Piotr Kocha\'nski\thanks{piotr@theta1.ifpan.edu.pl} $$ and
Krzysztof W\'odkiewicz\thanks{wodkiew@fuw.edu.pl} \\
Institute of Theoretical Physics, Warsaw University, \\ Warsaw 00-681, Poland}

\input{epsf}
\begin{document}
\maketitle
\begin{abstract}
The concept of intrinsic and operational observables in quantum mechanics
is introduced. It is argued that, in any realistic description of a quantum
measurement that includes a  detecting device, it is possible
to construct from the statistics of the recorded raw data a set of
operational quantities that correspond to the intrinsic quantum mechanical
observable. Using the concept of the propensity and the associated
operational positive operator
valued measure (POVM) a general description of the operational  algebra of
quantum mechanical observables is derived for a wide class of realistic
detection schemes. This general approach is illustrated by the example of an
operational Malus measurement of the spin phases and by an analysis of the
operational homodyne detection of the phase of an optical field with
a squeezed vacuum in the unused ports.
\end{abstract}

\section{Introduction}
In the standard formulation of quantum mechanics, the statistical outcomes of
an ideal measurement of an observable $\hat{A} |a\rangle = a |a\rangle$ are
described by the spectral measure [1]:

\begin{equation}
p_{\psi}(a) =|\langle a | \psi \rangle|^{2}
\end{equation}
where  $|\psi\rangle \in H$ is the
state vector of the measured system. The spectral measure contains all the
relevant statistical informations about the system, but it makes no reference
to the apparatus employed in the actual measurement. Because of this property
we shall refer to $\hat A$ as to an intrinsic quantum observable.

A realistic experiment necessarily involves additional degrees of freedom
[2] which eventually enable the experimenter to convert the raw data into a
operational  {\em propensity\/} density, 
$\Pr ({a})$ of a classical variable $a$ [3-4]. This propensity
depends on the state of the system and on all the devices used in a realistic
detection scheme. All these additional devices will be referred to as a
filter $\cal F$, that represents the experimental setup required for the
measurement of the observable $\hat{A}$. The measuring device is
described by the following positive and hermitian operator $\hat{\cal F}(a)$
that satisfies the relation:
\begin{equation}
\int da\ \hat{\cal F}(a) =1.
\end{equation}
In terms of this filter operator the propensity is 
\begin{equation}
\Pr ({a})=\langle\hat{\cal F}(a) \rangle .
\end{equation}
We see that in a realistic measurement the spectral decomposition of $\hat A$
is effectively replaced by a positive operator valued measure (POVM) $ da\
\hat{\cal F}(a)$ [5]. In view of the linear relation between the propensity
and the POVM, the operational statistical moments of the measured quantity
are:
\begin{equation}
\label{oodef}
\overline{{a}^n}
=\int\! {d}{a}\,{a}^n\Pr ({a})\ =\int\! {d}{a}\,{a}^n\langle\hat{\cal
F}(a)\rangle = \langle\hat A^{(n)}_{\cal F}\rangle 
\end{equation}
where 
\begin{equation}
\hat A^{(n)}_{\cal F}=\int\! {d}{a}\,{a}^n\hat{\cal F}(a)\,,
\end{equation}
defines a unique set of operational observables associated with a given POVM
for a given filter $\cal{F}$ [6].

As a rule, the algebraic properties of the $\hat A^{(n)}_{\cal F}$ operators
are quite different from those of the powers of $\hat A$. In particular, a
factorization is typically impossible, so that, for instance, 
$\hat A^{(2)}_{\cal F}$ does not equal $(\hat A^{(1)}_{\cal F})^2$. 

It is the purpose of this work to provide an explicit construction of the
POVM and the associated set of operational observables for two distinct
systems, both leading to an operational algebra of sinus and cosines
operators. The first system will be related to phases of the spin [7] probed
by the so called Malus filter [8], and the second system will be an optical
field probed by the so called  homodyne filter [9].

In both cases we shall derive operational operators corresponding to the
phases of the spin $s$ or of the optical field. These operational
observables will define an operational quantum trigonometry of the
corresponding phase measurements.

\section{Spin Operational Observables}
In this section we derive operational  operators of the spin phases and
describe a simple idealized experimental scheme leading to such operational
observables.
This experiment is based on the Malus law for spin. This law predicts that
 the transmission of a spin-$\frac{1}{2}$ through a measuring apparatus is
given
by $\cos^{2}\frac{ \alpha}{2}$, where $\alpha$ is the relative angle between
the orientation of the detected spin and the orientation of the Stern-Gerlach
polarizer. This
property can be generalized to a system with arbitrary spin $s$. We shall
assume that such a  system is described by spin coherent states
$|\Omega\rangle$, where the solid angle characterizes an arbitrary spin
orientation on a unit sphere. These spin $s$ coherent states are obtained by
a rotation of the maximum "down" spin state $|s,-s\rangle$ [10]:
\begin{equation}
|\Omega \rangle =\exp( \tau {\hat S}_{+} - 
\tau^{*} {\hat S}_{-})|s,-s\rangle ,
\end{equation}
where $\tau=\frac{1}{2}\theta e^{-i\phi}$ and   ${\hat S}_{\pm}$ are
the spin-$s$ ladder operators. 
The spin coherent states form an over complete set of states on the Bloch
sphere:
\begin{equation} 
\frac{2s+1}{4\pi} \int d \Omega\ 
 |\Omega\rangle \langle \Omega|= I. 
\end{equation}
Using these formulas, it is easy to obtain the
Malus  probability for a transmission of
such a  state through a Stern-Gerlach apparatus with orientation
$\Omega^{\prime}$. As a result one obtains:
\begin{equation}
p =|\langle \Omega|\Omega^{\prime}\rangle|^{2} = 
(\cos \frac{\alpha}{2})^{4s} .
\end{equation}
This quantum mechanical expression for 
the transmission function  provides a generalization of the 
spin-$\frac{1}{2}$ Malus Law  to the case of an arbitrary
spin $s$. A measurement leading to the Malus law can be easily constructed at
least in principle. Let assume that the Hilbert space of the system is
extended by a filtering device (another spin-$s$) initially in the "down"
spin state. A measurement is described by a dynamical process which generates
a correlation between the system being detected and the measuring filter. Due
to  the unitarity of the interaction with the filter, it is possible to
select the interaction parameters in such a way, that the wave function of
the combined system  evolves in the following way:
\begin{equation}
|s,-s\rangle_{\cal F} \otimes|\Omega\rangle\  \rightarrow \
|\Omega\rangle_{\cal F} \otimes|s,-s\rangle
\end{equation}
>From this relation is clear that a measurement of the filter orientation
leads to the spin  Malus law, which in the space of the detected spin is
equivalent to the following propensity:
\begin{equation}
\label{propensity-spin}
\Pr(\Omega^{\prime})=\frac{2s+1}{4\pi}
|\langle\Omega^{\prime}|\Omega\rangle|^{2}.
\end{equation}
This relation shows that the corresponding POVM is just:
\begin{equation}
\hat{\cal F}(\Omega)= \frac{2s+1}{4\pi}
|\Omega \rangle \langle \Omega | \ \ {\rm with} \ \ \int d\Omega \ \hat{\cal
F}(\Omega) =1.
\end{equation}
Having this simple picture of spin measurement we will look for
quantum operational observables connected with the Malus experiment.
There is a variety of operational operators that can be associated with
such phase measurements. For example the statistical moments of the azimuthal
orientation are given by:
\begin{equation}
\overline{\cos^{n}\theta}
=\int\! {d}{\Omega}\, \cos^{n}\theta \Pr ({\Omega})\  = \langle\hat
{\Theta}^{(n)}_{\cal F}\rangle 
\end{equation}
where the operational azimuthal cosine operators
\begin{equation}
\hat {\Theta}^{(n)}_{\cal F}=\int\! {d}{\Omega}\, \cos^{n}\theta\hat{\cal
F}(\Omega)\,,
\end{equation}
define a unique set of operational observables associated with this POVM.
All integrals in this expression can be calculated and we obtain:
\begin{equation}
\hat{\Theta}^{(n)}_{\cal F}={\rm F}(-n,s+\hat S_{3}+1,2s+2;2),
\end{equation}
where ${\rm F}(a,b,c;z)$ is a hypergeometric
function. The first two operational azimuthal cosine operators  are:
\begin{eqnarray}
\hat{\Theta}^{(1)}_{\cal F}&=&-\frac{1}{1+s}\hat{S}_{3}, \\
\hat{\Theta}^{(2)}_{\cal F}&=&\frac{2}{(1+s)(3+2s)}\hat{S}_{3}^{2}+
\hat{1}\frac{1}{3+2s}.
\end{eqnarray}
In the same way one can construct a corresponding set of operators describing
the operational properties of the polar coordinate of the spin system.
Statistical moments of the polar angle
\begin{equation}
\overline{\exp(in\varphi)}
=\int\! {d}{\Omega}\, \exp(in\varphi) \Pr ({\Omega})\  = \langle\hat
{E}^{(n)}_{\cal F}\rangle 
\end{equation}
lead to the following
operational set of polar phasors defined as
\begin{equation}
\hat{E}^{(n)}_{\cal F}= \int d\Omega\,
e^{in\varphi}\,\hat{{\cal F}}(\Omega)\, .
\end{equation}
We assume, that $n$ is a positive integer and that $\hat{E}^{(-n)}_{\cal
F}=\hat{E}^{(n)\:\dagger}$. Simple calculations give
\begin{equation}
\label{exp}
\hat{E}^{(n)}_{\cal F}=\hat{S}^{n}_{+}
\frac{\Gamma(s-\hat{S}_{3} +1-n/2)\Gamma(s+\hat{S}_{3}+1+n/2)}
{\Gamma(s+\hat{S}_{3}+n+1) \Gamma(s-\hat{S}_{3}+1)},
\end{equation}
with $\hat{E}^{(n)}_{\cal F}=0$ for $n>2s$ .
Two first moments are given by
\begin{eqnarray}
\hat{E}^{(1)}_{\cal F}&=&\hat{S}_{+}\frac{\Gamma(s-\hat{S}_{3}+1/2)
\Gamma(s+\hat{S}_{3}+3/2)}
{\Gamma(s+\hat{S}_{3}+2) \Gamma(s-\hat{S}_{3}+1)}, \\
\hat{E}^{(2)}_{\cal F}&=&
\hat{S}^{2}_{+}\frac{1}{(s+\hat{S}_{3}+2)(s-\hat{S}_{3})}.
\end{eqnarray}
So far we have derived operational operators associated only with polar
$\varphi$ and azimuthal $\theta$ directions of the spin. In the same way,
from the statistical properties of the spin propensity, it is possible to
derive operational spin operators. These operators correspond to Malus
measurements of unit directions with
a filter defined by a spin coherent state POVM. 
We can parameterize the three spin coordinates by a solid angle on a
unit sphere in the following way
\begin{eqnarray}
\hat{S}_{1}\longrightarrow\cos\varphi\sin\theta,\,\,\,\,
\hat{S}_{2}\longrightarrow\sin\varphi\sin\theta,\,\,\,\,
\hat{S}_{3}\longrightarrow -\cos\theta.
\end{eqnarray}
The corresponding  spin operational operators may be naturally defined as
follows
\begin{equation}
\nonumber
\hat{\Sigma}^{(n)}_{i}=\int d\Omega\,(n_{i})^{n}\,
\hat{{\cal F}}(\Omega)\,,\,\,\,\,\, i=1,2,3\,,
\end{equation}
where $\vec{n}=(\cos\varphi\sin\theta,\,\sin\varphi\sin\theta,\,-\cos\theta)$
is a unit vector.

In further discussion we concentrate only on the first two  operators from
the whole operational spin algebra
\begin{eqnarray}
\hat{\Sigma}^{(1)}_{i}&=&\frac{1}{1+s}\hat{S}_{i}, \\
\hat{\Sigma}^{(2)}_{i}&=&\frac{2}{(1+s)(3+2s)}\hat{S}_{i}^{2}+
\hat{1}\frac{1}{3+2s}.
\end{eqnarray}
Operational spin observables are proportional to the intrinsic spin
operators, however they are modified due to the noise imposed by
the measuring apparatus forming the Malus analyzer.

\section{Squeezed Quantum Trigonometry}

As a second example of a possible application of the  presented
formalism we give a theoretical description and generalization of the
experiments recently
performed by Noh, Foug\`eres and Mandel [9].
The authors have used an eight-port homodyne detector (NFM apparatus)
in order to measure the relative phase between two classical or quantum
light fields.
In such an experiment we measure the difference of the photon counts on two
pairs of detectors. This quantity is either related
to the sine and cosine of the phase difference of two
classical electromagnetic fields (classical case) or may be used to
define the set of operational operators associated with an arbitrary
classical function of the phase (quantum case).
Particularly, we can find quantum analogs of trigonometric
functions and their powers obtaining the so called ``quantum trigonometry''
[11].

Below we derive such a ``quantum trigonometry'' for modified NFM
apparatus. We will assume here, that the additional noise 
coming through the two free unused ports of the NFM experimental device is
described by  squeezed vacuum state (in the original NFM experiment this
field was  a vacuum state). Because of this the resulting operational
algebra will be represent by a ``squeezed quantum trigonometry''.

In our case the propensity $\Pr(\varphi;s,\phi)=\Pr(\varphi+2\pi;s,\phi)$
is a periodic function of the phase, normalized to unity in the following way
\begin{equation}
\label{normalization}
\int \frac{d\varphi}{2\pi}\Pr(\varphi;s,\phi)=1\,.
\end{equation}
By $s$ and $\phi$ we denoted the amplitude and the phase of the squeezed
vacuum.
\begin{figure}[t]
\begin{center}
\begin{tabular}{rcl}
\put(84,45){$I$}
\put(225,80){$\varphi$}
\leavevmode  \epsfxsize=9.0cm \epsffile{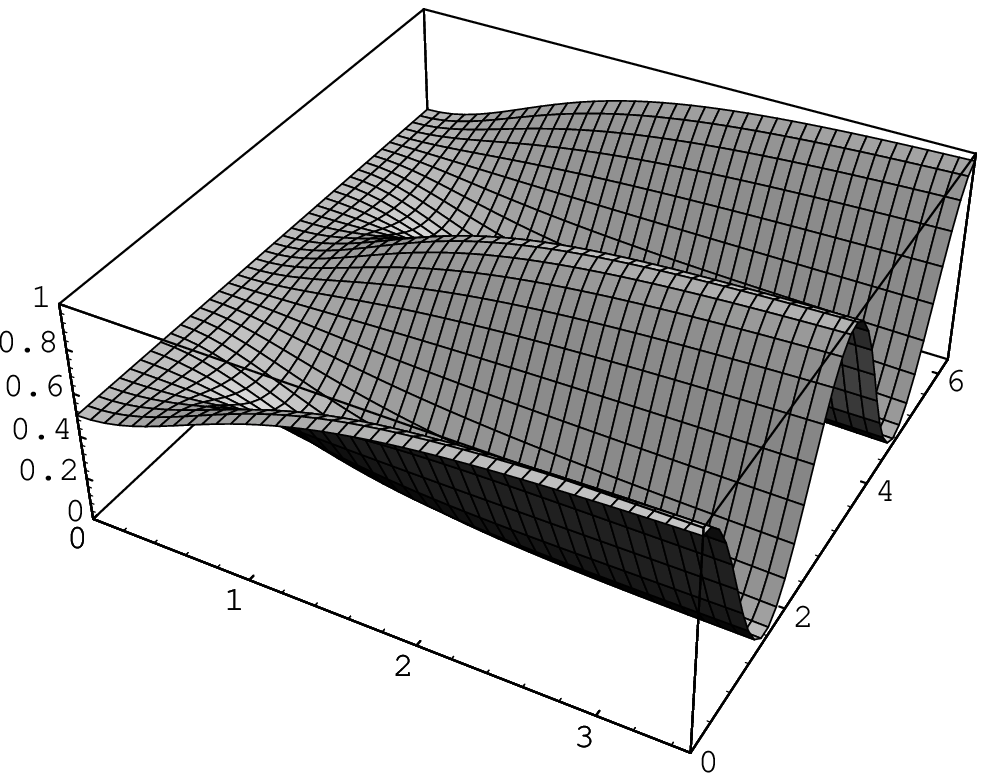} \\
\end{tabular}
\end{center}
\centerline{Fig. 1:  Wigner function of operational operator
$\hat{C}^{(2)}(s=0.5,\phi=\pi/2)$.}
\label{c1}
\end{figure}

\begin{figure}[t]
\begin{center}
\begin{tabular}{rcl}
\put(84,45){$I$}
\put(225,80){$\varphi$}
\leavevmode  \epsfxsize=9.0cm \epsffile{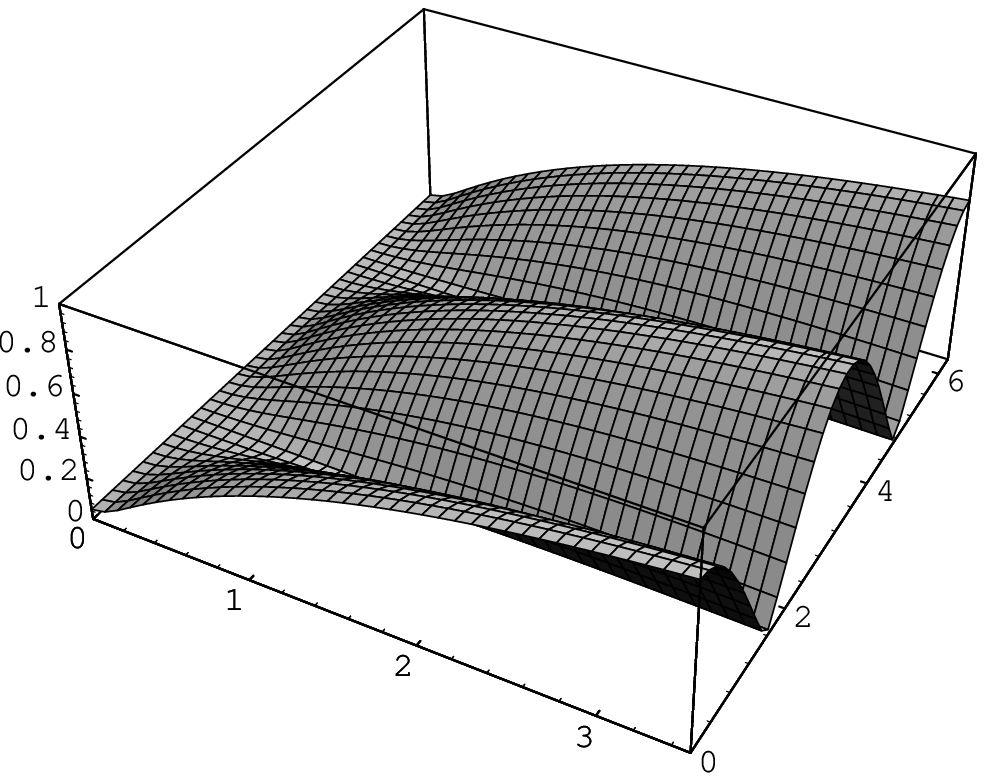} \\
\end{tabular}
\end{center}
\centerline{Fig. 2  Wigner function of operational operator  
$\hat{C}^{(2)}(s=1.5,\phi=0)$.}
\label{c2}
\end{figure}
In accordance with the general scheme (\ref{oodef}) the phasor operators
corresponding to an operational squeezed quantum trigonometry are defined
as follows
\begin{equation}
\label{meanv-phasor-def}
\overline{e^{in\varphi}}\equiv\langle \hat{E}^{(n)}(s,\phi)\rangle,
\;\;\; n=\pm1, \pm2, \ldots\,.
\end{equation}
As in [11], we assume from the beginning, that the local
oscillator (reference field) is a strong coherent laser field and therefore
we are allowed to neglect the noise of the input field  mixed
by the beam splitter with the reference signal.

We start with the formula obtained in [11]
\begin{equation}
\label{phasor}
\hat{E}^{(n)}(s,\phi)=
{\rm Tr}_{v}\left\{\frac{(\hat{b}+\hat{v}^{\dagger})^{n}}
{\left((\hat{b}+\hat{v}^{\dagger})
(\hat{b}^{\dagger}+\hat{v})\right)^{\frac{n}{2}}}
\hat{S}(s,\phi)|0_{v}\rangle\langle 0_{v}|\hat{S}^{\dagger}(s,\phi)
\right\}\,,
\end{equation}
and change it for our purpose using a unitary operator $\hat{S}(s,\phi)$,
which generates the squeezed vacuum state $|s\rangle\langle s|$ from the
vacuum state $|0\rangle\langle 0|$ [12].
The bosonic creation and annihilation  operators
 $\hat{v}^{\dagger}$, $\hat{v}$ represent an additional degree of freedom
associated with the squeezed vacuum input at the beam splitter,
by  $\hat{b},\;\hat{b}^{\dagger}$ we denote the creation and the annihilation
operators of the signal field. Because of the fact, that the trace
${\rm Tr}_v$ in (\ref{phasor}) is invariant under any unitary transformation,
variables connected with the local oscillator can be removed and the only
place in (\ref{phasor}), where the reference field contributes is the
operator $\hat{S}(s,\phi)$ with $\phi$ shifted by an unimportant phase that
we shall ignore.

Using various properties of coherent and squeezed states
[12],  the trace in (\ref{phasor}) can
be calculated and we have
\begin{equation}
\label{phasor-int}
\hat{E}^{(n)}(s,\phi)=
\int\frac{d^{2}w}{\pi}\frac{w^{n}}{(w^{\ast}w)^{\frac{n}{2}}}
|w,s\rangle\langle w,s|,
\end{equation}
where $|w,s\rangle ={\hat D}(w) \hat{S}(s,\phi) |0\rangle$ is the squeezed
coherent state with amplitude $w$ and squeezed parameters $s$ and $\phi$.
According to our terminology
$\hat{{\cal F}}(s,\phi)=|w,s\rangle\langle w,s|$
is the positive valued operator measure (POVM) associated with
the described experimental scheme.

An exact formula for the phasor may be derived straight 
from (\ref{phasor-int}). Recalling the unity decomposition for
the (squeezed) coherent states we obtain
\begin{equation}
\label{phasor-formula}
\hat{E}^{(n)}(s,\phi)=\hat{S}(s,-\phi)
\vdots\frac{(\hat{b}\cosh s-\hat{b}^{\dagger}e^{i\phi}\sinh s)^{n}}
{\left[(\hat{b}\cosh s-b^{\dagger}e^{i\phi}\sinh s)
(\hat{b}^{\dagger}\cosh s-\hat{b}e^{-i\phi}\sinh s)\right]^{\frac{n}{2}}}
\vdots\hat{S}^{\dagger}(s,-\phi).
\end{equation}
When the squeezing parameter $s$ tends to zero the above formula reduces to
the results obtained in the reference [11].

The propensity density can be  simply derived from (\ref{phasor-int}). If
we set $ w= \sqrt{I} \exp(i\varphi)$, we obtain the propensity in the form of
the following marginal integration
\begin{equation}
\Pr(\varphi;s,\phi)=\int_{0}^{\infty}dI\ \langle w,s|\hat{\rho}|w,s\rangle.
\end{equation}
It's clear, that the propensity, contrary to the quantum mechanical 
probability density, depends  on the experimental device. In fact,
for each value of the squeezing parameter $s$ we obtain a 
different propensity $\Pr(\varphi;s,\phi)$ and a different phasor basis,
even though the probe field remains unchanged.

As it is easy to see, phasors (\ref{phasor-formula}) are not Hermitian
operators so they cannot correspond to observable quantities.
Nevertheless, using phasor basis (\ref{phasor-formula})
it is possible to define naturally ``trigonometric operators'', whose
mean values can be measured in a real experiment --- for $s=0$ they
have been actually measured by Noh et al. [9].
For example, two first ``cosine'' operators are defined in the following way
\begin{eqnarray}
\label{trigfun-def}
\hat{C}^{(1)}\equiv\frac{1}{2}(\hat{E}^{(1)}+\hat{E}^{(-1)}), \nonumber \\
\label{trigfun}
\hat{C}^{(2)}\equiv\frac{1}{2}+\frac{1}{4}(\hat{E}^{(2)}+\hat{E}^{(-2)}).
\end{eqnarray}
In a similar manner we can find moments of ``sine'' operators
or, if it's needed, of any periodic function of the phase, provided
we know its Fourier decomposition. Replacing the Fourier components
 $\exp{(in\varphi)}$ by the corresponding $n$-th phasors, we construct  in
such a way the operational operator corresponding to an arbitrary function of
the phase.

In order to investigate the properties of the phasors we evaluate
(numerically) the corresponding Wigner functions of these operators.
Examples of such Wigner functions  are presented in 
fig. 1 and fig. 2.

First we notice, that, according to the terminology introduced in [13],
the phasors have a  proper classical limit. If the incoming  intensity of the
signal field tends to infinity, the Wigner functions of the operational
phasors reproduce a classical trigonometry.
This limit can be seen in the fig. 1. and in fig. 2.
Comparing both figures, we  observe, that an increase of $s$ causes 
a reduction of the phasors amplitude.  
As it might have been expected, the dependence on
the squeezing parameters is gone in the classical limit.

In the limit of very small $I$, and with the squeezed phase $\phi$ equal to
$0$ or $\pi$ we have
\begin{equation}
C^{(1)}_{W}(s,\phi)\stackrel{I\rightarrow 0}{=}
\sqrt{I}{\cal A}_{0,\pi}(s)\cos\theta
\end{equation}
where ${\cal A}_{0,\pi}$ are two amplitudes of the Wigner function,
that depend only on the squeezing parameter $s$ and $\phi=0$ or $\phi=\pi$.
This result shows that the amplitude of the  cosine Wigner function is
literally "squeezed".
For arbitrary values of $\phi$ the separation of the cosine Wigner function
into an amplitude, and a purely angle dependent part is no longer possible,
but a clear squeezing of the amplitude is also observed [14]. For $I=0$ the
Wigner cosine function is zero, which is in agreement with the property that
the phase of a light field in the vacuum state is randomly distributed.

Similarly we can find an asymptotic expression for $C^{(2)}_{W}(s,\phi)$.
In the limit of small $I$
\begin{equation}
C^{(2)}_{W}(s,\phi)=\frac{1}{2}(1-c(s,\phi)),
\end{equation}
where $c(s,\phi)$ is an $I$-independent function of the squeezed parameters.
It's easy to check, that $c(s,\phi)$ tends either to unity ($\phi=0$)
or to minus unity ($\phi=\pi$).
As a result, for small intensities $C^{(2)}_{W}(s,\phi)$
becomes zero ($\phi=0$) or one ($\phi=\pi$). For
$\phi=\pi/2,\,3\pi/2$ $c(s,\phi)=0$ and
$\lim_{I\rightarrow 0}C^{(2)}_{W}(s,\phi)=1/2$.
For small $I$ the squeezing
influences the system very strongly. If the squeezed phase $\phi$ equals
to $\pi/2$, the phasor's $\hat{C}^{(2)}$ Wigner 
function tends $1/2$
(fig. 1), whereas for $\phi=0$ it takes values near  zero
(fig. 2). Such a dramatic change of the cosine quadratures occurs because
in the limit of small $I$,  purely quantum effects of the squeezed vacuum are
important. The squeezing allows one of the quadratures to be reduced below
the vacuum level represented by a uniformly distributed random phase. 
The uniform distribution of the phase corresponding to the vacuum state
leads to an operational quadrature  $\frac{1}{2}$.
For a squeezed vacuum, this uniform random-phase distribution is modified
[15] and a significant change of the operational quadrature is possible.
In fact fluctuations below $\frac{1}{2}$ in the Wigner
function exhibit the quantum nature of the squeezed vacuum. 

Another interesting observation can be made if we look at the phasor's
squeezed coherent states POVM decomposition (\ref{phasor-int}) and
recall the fact, that the  Glauber $P$-representation of a  
squeezed  coherent state does not exist. This property is related to the
dynamical ordering of the creation and 
annihilation operators, induced by the measuring device.
For  the modified NFM apparatus, with a squeezed
vacuum in the unused port, the antinormal ordering of operational phasor is
impossible to achieve.

\section{Acknowledgment}
This work has been partially supported by the Polish KBN grant 2 PO3B 006 11.

\section*{References}
\begin{description}
\item[[1]]
J. von Neumann, {\sl Mathematische Grundlagen der Quantenmechanik},
Springer--Verlag, Berlin 1932; 

\item[[2]]
See, for example, {\sl Quantum Theory of Measurement}, edited by
J.~H.~Wheeler, W.~H.~Zurek, Princeton University Press, Princeton 1983;

\item[[3]]
K. W\'odkiewicz, Phys.Rev. Lett. {\bf 52}, 1064 (1984);
\item[[4]]
K. W\'odkiewicz, Phys. Lett. A {\bf 115}, 304 (1986);

\item[[5]]
See, for example, P. Busch, P. J. Lahti, and P. Mittelstaedt
{\sl The Quantum Theory of Measurement}, Springer--Verlag,
Berlin 1991; 

\item[[6]]
B.--G. Englert, K. W\'odkiewicz, Phys. Rev. A{\bf 51}, R2661 (1995);

\item[[7]]
A. Vourdas, Phys. Rev. A {\bf 41}, 1653 (1990); 

\item[[8]]
K. W\'odkiewicz, Phys. Rev. A {\bf 51}, 2785 (1995); 

\item[[9]]
J.W. Noh, A. Foug\`eres, L. Mandel, Phys. Rev. Lett. {\bf 71}, 2579 (1993); 

\item[[10]]
F.T. Arecchi, E. Courtens, R. Gilmore, H. Thomas,
      Phys. Rev. A {\bf 6}, 2211 (1972); 

\item[[11]]
B.--G. Englert, K. W\'odkiewicz, P. Riegler,
      Phys. Rev. A {\bf 52}, 1704 (1995);  

\item[[12]]
H. Yuen, Phys. Rev. A {\bf 13}, 2226 (1976);

\item[[13]]
J. Bergou, B.--G. Englert, Ann. Phys. (NY) {\bf 209}, 479 (1991); 

\item[[14]]
P. Kocha\'nski and K. W\'odkiewicz, to be published (1996);     
\item[[14]]
D. Burak and K. W\'odkiewicz, Phys. Rev A {\bf 46}, 2774 (1992);
\end{description}
\end{document}